\documentclass[11pt]{article}
\usepackage{epsfig}
\renewcommand{\baselinestretch}{1.2}

\graphicspath{{images/}}
\usepackage{graphicx}
\usepackage{oldgerm}
\usepackage{float}
\usepackage{verbatim}
\usepackage{hyperref} 
\usepackage{subfig}
\usepackage{float}
\usepackage{epsfig}
\usepackage{epstopdf}
\def\be{\begin{equation}}
\def\ee{\end{equation}}

\newsavebox{\PSLASH}
\sbox{\PSLASH}{$p$\hspace{-1.8mm}/}

\usepackage{setspace}

\begin{document}

\bibliographystyle{plos2009}


%
%
%

\begin{flushleft}
{\Large
\textbf{The effect of temporal pattern of injury on disability in learning networks}
}
\\
‍Mohammadkarim Saeedghalati$^{1,\ast}$, 
Abdolhossein Abbassian$^{1}$, 
\\
\bf{1} Biomath section, Mathematics department, IPM, Tehran, Iran
\\
$\ast$ E-mail: arsham@gmail.com
\end{flushleft}

\section*{Abstract}
How networks endure damage is a central issue in neural network research. This includes temporal as well as spatial pattern of damage. Here, based on some very simple models we study the difference between a slow-growing and acute damage and the relation between the size and rate of injury. Our result shows that in both a three-layer and a homeostasis model a slow-growing damage has a decreasing effect on network disability as compared with a fast growing one. This finding is in accord with clinical reports where the state of patients before and after the operation for slow-growing injuries is much better that those patients with acute injuries.

\textit{Keywords}: Temporal pattern of injury, Learning networks, Homeostasis, Neural networks.

\section{Introduction}
The performance of networks in the case of damage is a central issue in biological as well as non-biological networks \cite{Nawrocki2011,Hinton2000}. Intuitively the degree of disability following a damage must be correlated with the size of damage. However, different networks such as brain network, small world, scale free and random networks show different responses to spatial pattern of damage \cite{Alstott2009,Fortney2007,Vazquez2002,Albet2001,Callaway2000,Cohen2000}.Although there are many works ralated to network recovery \cite{Lee2011,Wang2011,Murre2003}, dependence of network recovery on network architecture in the case of spatial patterns of damage remains to be further explored. Another basic question is to consider the temporal patterns of damage. For example, a slow growing damage may lead to better recovery whereas a fast growing damage may lead to a recovery failure given a maximum size of spatial damage \cite{Desmurget2007,Duffau2002,Duffau2003,Varona2010}. There are experimental data on rat, cat and also monkey that shows a delay time between brain lesions has a strong effect on deficits that animals bear after operation \cite{STEWART1951,Meyer1958,Glick1972,Rosen1971,Adametz1959,Finger1971,Patrissi1975}.In a more clinical setting this proves to be a crucial aspect of how the brain recovers  from strokes or some lesion growing tumors. The biological basis of recovery after stroke and how the brain reorganizes itself, for example, is still largely unknown \cite{Calautti2003,Calautti2003a} but certainly it is possible that the degree and speed of recovery  varies considerably for different lesion locations and depends on structural alterations taking place in the spared brain tissue \cite{STEWART1951,Meyer1958,Glick1972}. Recently there has  been some attempts using simple recurrent networks to show the effect of brain remapping and how it contributes to neuronal homeostasis following lesions\cite{Butz2008,Butz2009}.These findings, however, relate more to the spatial pattern of brain lesions and needs to be extended to what may be called a spatio-temporal pattern of brain injury. Here, a distinction has to be made between an acute stroke where there is a sudden damage inflicted to the brain and a slow growing damage such as a low-grade gliomas\cite{Desmurget2007,Duffau2002,Duffau2003,Varona2010}.  Based on many experimental reports the slow growing injuries have a much better chance of being recovered than the injuries caused by acute lesions\cite{STEWART1951,Meyer1958,Glick1972,Rosen1971,Adametz1959,Finger1971,Patrissi1975}. As one might say if neurons can will, recovery comes as no surprise\footnote{we thank neurosurgeon Dr. K. Parsa for bringing this remark to our attention}.
In the present paper two simple networks, a three-layer and a homeostasis model with different temporal pattern of damages are studied. As will be shown a gradual injury is indeed resisted by these networks providing the lesions inflicted on the networks do not exceed a critical size. Our main finding is that a gradual injury will decrease the maximum amount of disability that the network may tolerate before it brakes down with no chance of recovery.

\section{Three layer model}
\label{Three layer model}
\subsection{The model}
\begin{figure}[ht]
\begin{center}
\centerline{\includegraphics[height=7cm]{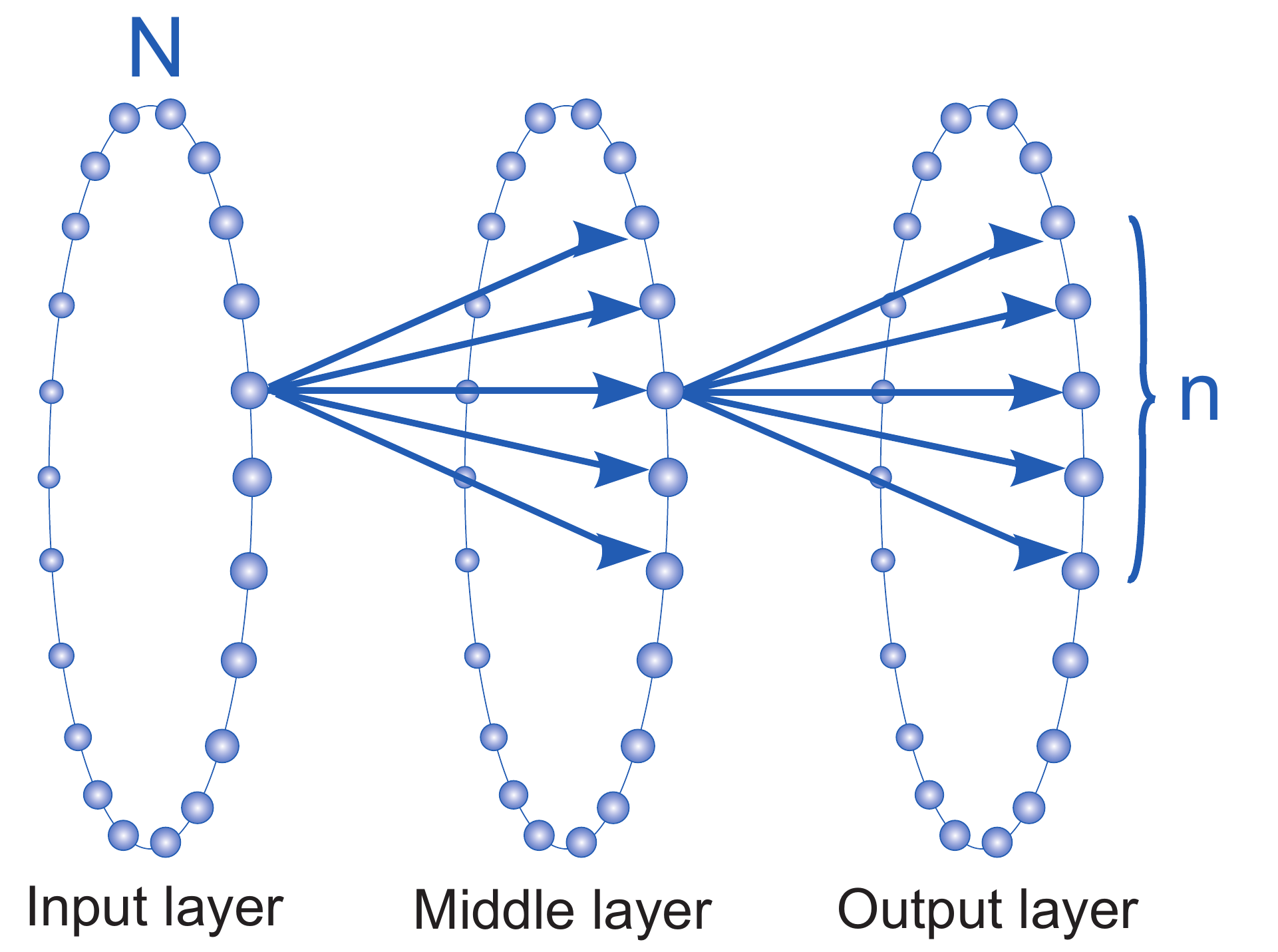}}
\protect
\caption{\footnotesize
\label{3layernet}
Three layer model network topology: each one of $N$ nodes from input layer connects to $n$ neighbor nodes in the process layer which are symmetrically asides the one which is in front of the node in the input layer. The same connections hold from process layer to the output layer. There isn't any connections between nodes of each layer but there is feed-forward connection from previous layer to the next one.
   }
\end{center}
\end{figure}

In this model the nodes are binary with state $1$ means active and state $0$ inactive. The network learns the correct answer for each input. A measure of disability, the Hamming distance, is defined as the distance between the desired and the actual output. The network consists of three rings of nodes each containing $N$ nodes. There is no connections between nodes in each layer but there is feed-forward connection from the previous to the next layer. Each node $i$  is  connected to a group of $n$ neighbor nodes in the next layer (see figure\ref{3layernet}).

The fraction $\frac{n}{N}$ measures network completeness in the sense that if we remove a group of $m$ neighbor nodes in the middle layer and $m > n$, the route from input layer to output layer for some nodes may be cut. For $m<n$, although the network may be initially disabled has the potential to recover. There is a sigmoid activation function for the middle layer and nodes in the output layer are binary threshold neurons. In each trial one input is shown to the first layer and the connection weights are changed according to a Hebbian-like learning rule as long as a difference exists between the desired and the actual output. 

In the pre-lesion phase we start with random binary patterns in  the  input and output layer and give enough time for the system to learn. In the lesion phase we  kill  some  nodes immediately   or   gradually   in the middle layer and wait for recovery of the system. At each time step we calculate hamming distance in the output layer as a measure to quantify network disability.
There are two approaches in the lesion phase. In the first one we remove the nodes gradually one by one until the total number of $n_l$ nodes are removed. In the other approach according to the resection experiments \cite{STEWART1951,Meyer1958,Glick1972,Rosen1971,Adametz1959,Finger1971,Patrissi1975} the total number removed nodes, $n_l$, is divided in to $n_p$ equal size packages where at each time step a package containing $\frac{n_l}{n_p}$ nodes is removed. 
The simulation details and parameter values for this part can be find in the Appendix \ref{3layerapp}.
\subsection{Discussion on the learning rate, $\eta$}
\label{modif-eta}
\begin{figure}[ht]
\begin{center}
\centerline{\includegraphics[height=6cm]{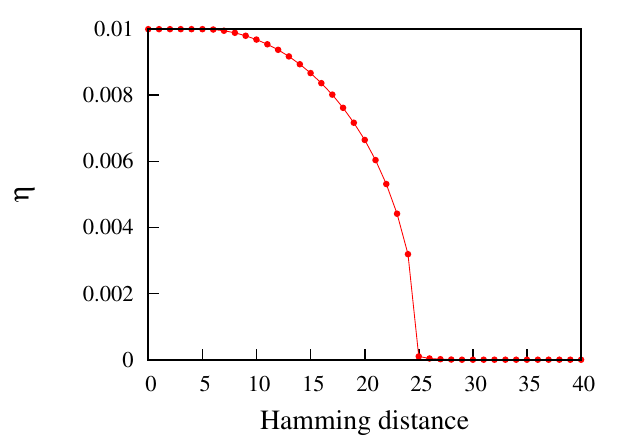}}
\caption{\footnotesize
\label{neweta}
A more realistic relation between $\eta$ and network disability.  The $\eta$ will decrease with the size of network disability as the injury may ruin the learning part of the system too. There   is   a  threshold size that  for injuries  larger  than that, the system could not be recovered anymore.
   }
\end{center}
\end{figure}

The learning rate was assume constant in \ref{Three layer model}. In a more realistic network $\eta$ is held to be a function of network disability. We   suppose   that   there  is  a  threshold size  of injuries above which the system may not be recovered.  The $\eta$ will decrease with the size of network disability as the injury may ruin the learning part of the system too. We suppose that the functionality of $\eta$ with network disability behaves as in figure \ref{neweta}.
In the homeostasis model described in section \ref{Homeostasis model} the learning rate is also proportional to network disability.

\subsection{Three layer model results}

\subsubsection{Three layer model with fixed $\eta$}

\begin{figure}[ht]
\begin{center}
\begin{tabular}{cc}
\subfloat[]{\includegraphics[width=0.4\textwidth]{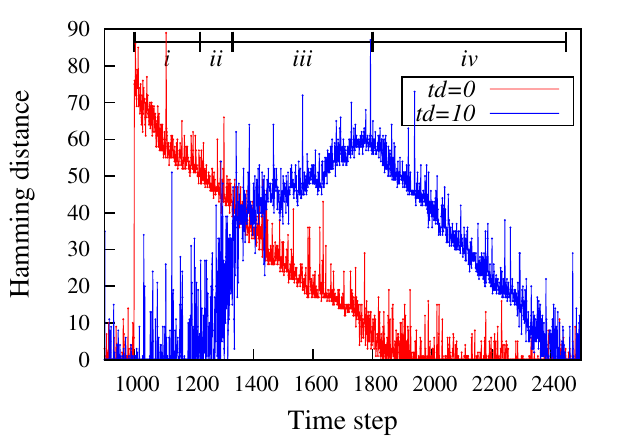}\label{3layerres-f-one}} 
   & \subfloat[]{\includegraphics[width=0.4\textwidth]{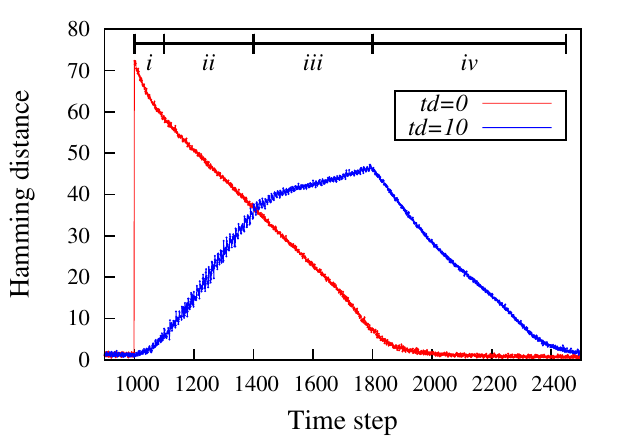}\label{3layerres-f-200}}\\
\subfloat[]{\includegraphics[width=0.4\textwidth]{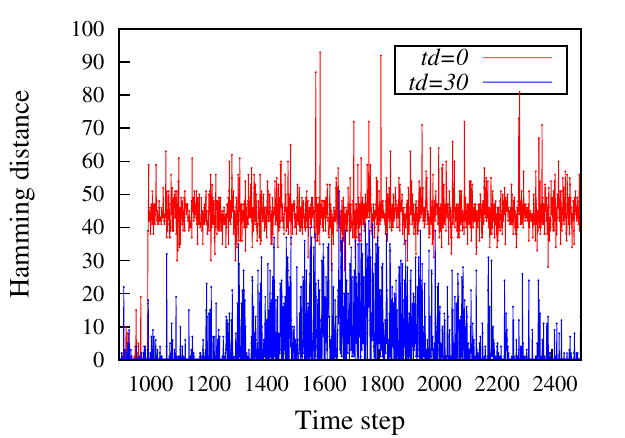}\label{3layerres-m-one}} 
   & \subfloat[]{\includegraphics[width=0.4\textwidth]{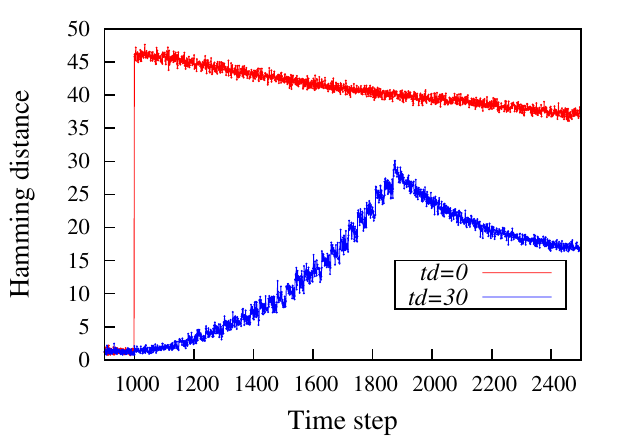}\label{3layerres-m-200}}\\
\end{tabular}
\end{center}
\caption{\footnotesize
\label{3layerres}
(a):The amount of network disability (Hamming distance) is shown for sudden and gradual node destruction. In the red curve where we remove the nodes suddenly the amount of network disability is more than in the blue one with gradual node removing. The existence of a critical injury size could be observed from the blue curve. During removing of the first $20$ nodes the network shows a small disability (i). After which the hamming distance suddenly increases and the network shows a severe disability (ii). Removing more nodes from $1300$ to $1800$ doesn't change network disability a lot (iii). After removing last node, network start to recover linearly like the blue curve (iv).
(b): All the the parameters in this diagram is equal to the figure \ref{3layerres-f-one}, but this one is the average of $200$ different runs. The behavior in the lesion phase is the same as in figure \ref{3layerres-f-one}.
(c): simulation with modified $\eta$ for injury size of $30$ for a single simulation. Red and blue curves are correspond to delay time (td) of $0$ and $30$ respectively.  
(d): simulation with modified $\eta$. For this curves we take the average of $200$ runs with different initial conditions. The other parameters are the same as in figure \ref{3layerres-m-one}.
   }\label{Mfigure}
\end{figure}

The results for immediate and gradual injury for delay time $"td"$ of $10$ and total injury size of $80$ is shown in figure \ref{3layerres-f-one}. In this diagram vertical axis indicates hamming distance as a parameter that shows the amount of network disability and horizontal axis is time. According to our three layer model, recovery/learning rate is a constant function of disability size. This is confirmed by the linear behavior of recovery shown in the red curve. In this diagram injury starts at time $1000$. The red curve shows a sudden and large network disability which linearly recovers over time. In the blue curve although we start injury at time $1000$,the nodes are removed gradually each $10$ time step. Up to time $1200$ the network somehow resists sudden huge disability but after $20$ nodes being removed, the network suddenly shows a huge disability. From $1300$ to $1800$ when the last of $80$ nodes is removed, there isn't a great change in network disability. At time $1800$ the network starts to 
recover linearly and behave just like the red curve. There are two important things about this diagram. First, the maximum disability decreases as the nodes are gradually removed with increasing delay time. 

There is however a new finding, to the effect that as we remove the nodes gradually the network disability doesn't increase gradually: Up to removing a specific number of nodes the network shows resistance but then suddenly changes its behavior and shows a great disability after which removing more nodes results in slower increase in disability.

 This behavior points to a possible critical size of injury. To study that these results not just hold for a specific initialization we simulate the three layer model with the same parameters but for $200$ different initializations. In the figure \ref{3layerres-f-200} the average of these simulations is shown. In this diagram the effects which we saw before is hold and also because of the small fluctuations of this diagram we observe new behavior in the recovery phase.

\subsubsection{Three layer model with modified $\eta$}
For better observation of this critical injury size effect we use the specific relation between learning rate $\eta$, and disability size described in  section \ref{modif-eta} (Also, see figure \ref{neweta}). With this dependency of $\eta$ and hamming distance(disability size) we could capture the state of severe disability. In this study we set this critical hamming distance to $25$. For hamming distances larger than $25$ the learning rate $\eta$, is close to zero and therefore no learning is observed.
With this newly defined $\eta$, we simulate three layer model for injury size of $30$ with this newly defined $\eta$. Figure \ref{3layerres-m-one} show two single runs, one for immediate and  one for delay time of $30$ between removing nodes. In the first diagram the Maximum amount of disability (MAoD) never reaches $25$ so the system could recover but in the next diagram where injury is immediate, At the begining the MAoD became larger than $25$ so the recovery rate $\eta$ drops down and system couldn't recover. In the figure \ref{3layerres-m-200} the same result for the average of $200$ different initializations is shown.

\subsubsection{Three layer model - Different kinds of injury}
In the previous sections the behavior of three layer model for gradual injury has been shown briefly. Here according to injury and recovery models four kind of sub-models has been developed:\\
Type 1: removing each node gradually with a delay between removing them when we use a fixed $\eta$ as a learning rate.\\
Type 2: removing each node gradually with a delay between removing them when we use a modified $\eta$ as a learning rate.\\
Type 3: removing nodes in packages gradually with a delay between removing them when we use a fixed $\eta$ as a learning rate.\\
Type 4: removing nodes in packages gradually with a delay between removing them when we use a modified $\eta$ as a learning rate.\\
In each sub-model the effect of delay has been studied. 
\subsubsection*{Type 1: No packages - Fixed $\eta$}
\begin{figure}[ht]
\begin{center}
\begin{tabular}{cc}
     \subfloat[]{\includegraphics[trim = 8mm 65mm 10mm 69mm, clip, width=0.5\textwidth]{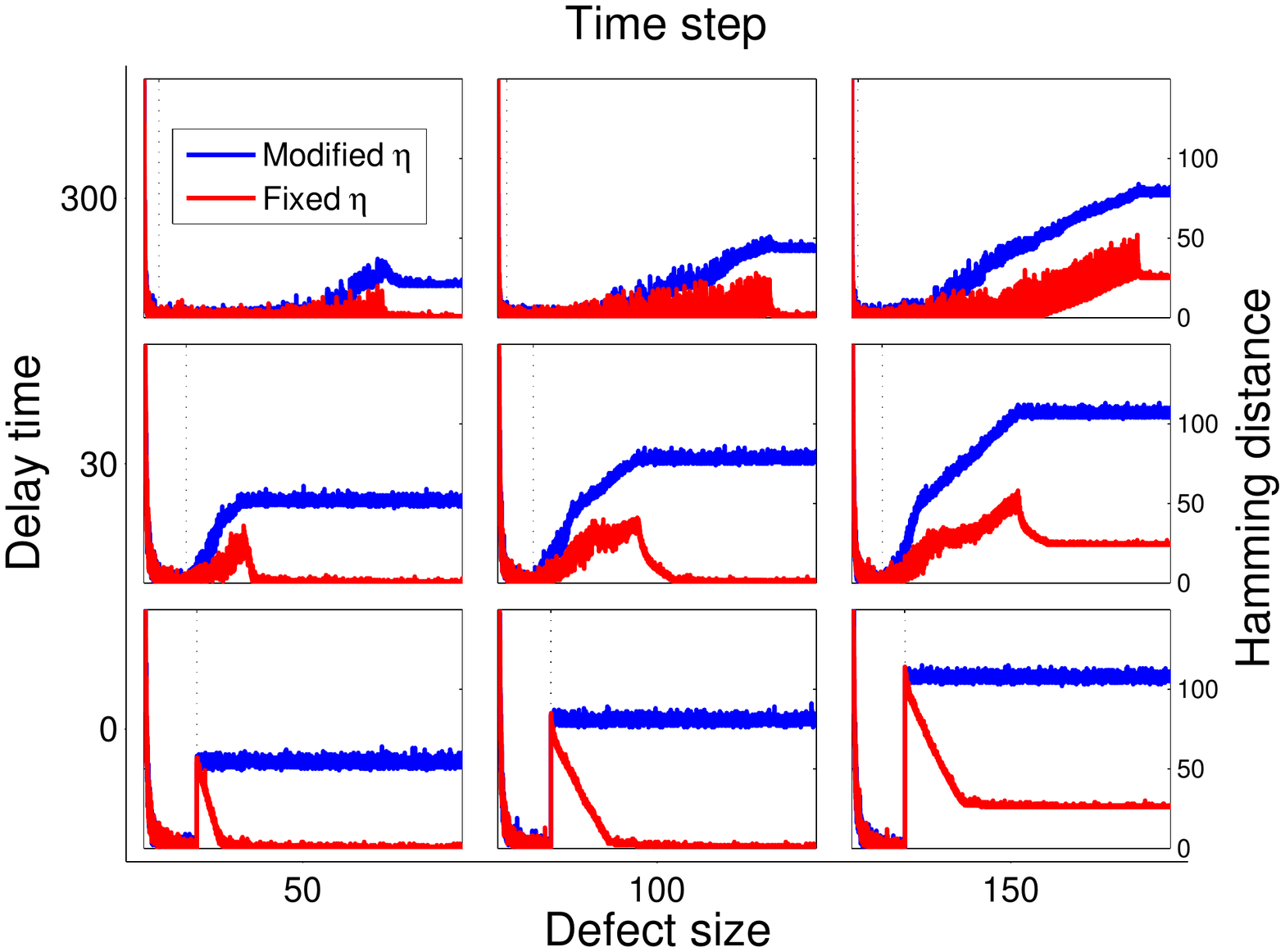}\label{l-gb-s-t-res}} 
   & \subfloat[]{\includegraphics[trim = 8mm 65mm 10mm 69mm, clip, width=0.5\textwidth]{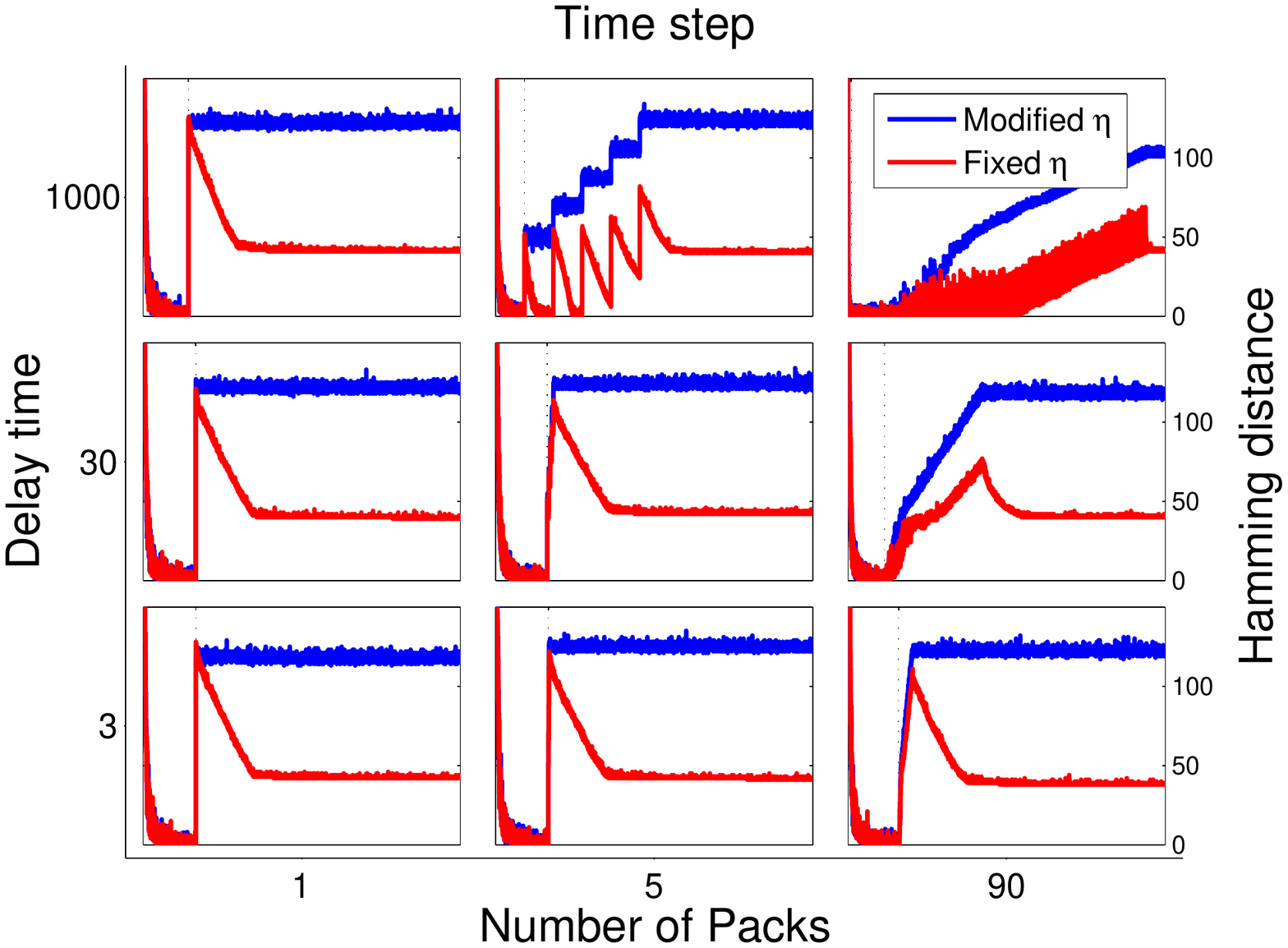}\label{l-b-s-t-res}} 
\end{tabular}
\end{center}
\caption{\footnotesize
\label{land-res}
(a): The effect of injury size and delay between removing the nodes had been shown. In each sub plot the amount of network disability as a function of time is shown for fixed (red) and modified $\eta$ (blue). For larger delay times, even for large injury sizes the network shows very small disability. vertical grid line in each subplot shows time step $1000$ which is the injury start time.
(b): The effect of number of packages and delay between removing them had been shown. In each sub plot the amount of network disability as a function of time is shown for fixed (red) and modified $\eta$ (blue). Increasing the number of packages and delay, decrease the MAoD. This diagram is an average of nine runs with different initial conditions. 
   }\label{Mfigureland}
\end{figure}

For the maximum amount of 180 nodes and delays in the range of $0$, sudden injury, to $1000$ time steps, different systems had been simulated for $9$ different initialization. A small subset of these simulations had been shown in red in the figure \ref{l-gb-s-t-res}.

\begin{figure}[ht]
\begin{center}
\begin{tabular}{cc}
\subfloat[]{\includegraphics[trim = 5mm 65mm 15mm 65mm, clip, width=0.4\textwidth]{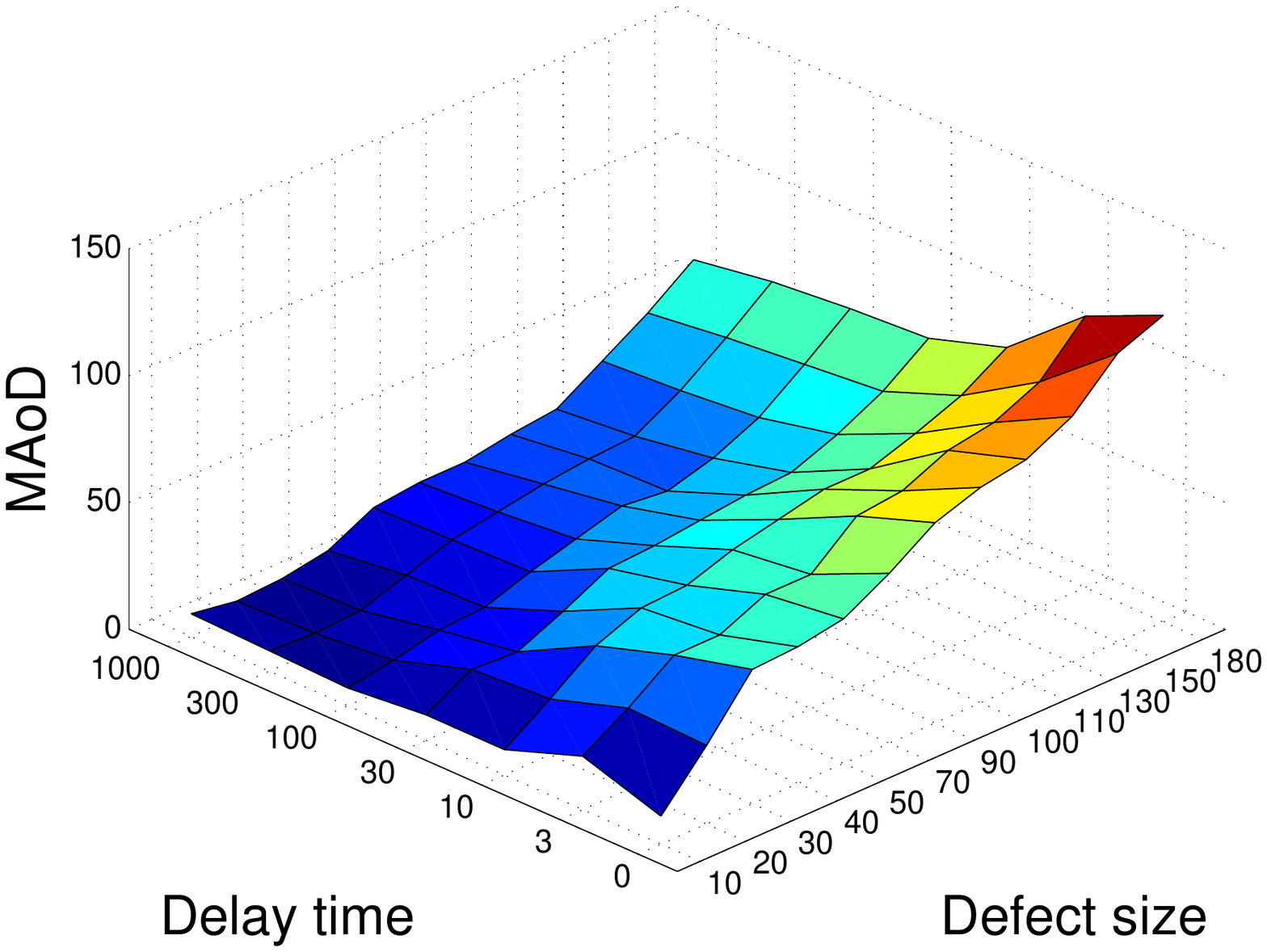}\label{l-gb-s-3d-s}} 
   & \subfloat[]{\includegraphics[trim = 5mm 65mm 15mm 65mm, clip,width=0.4\textwidth]{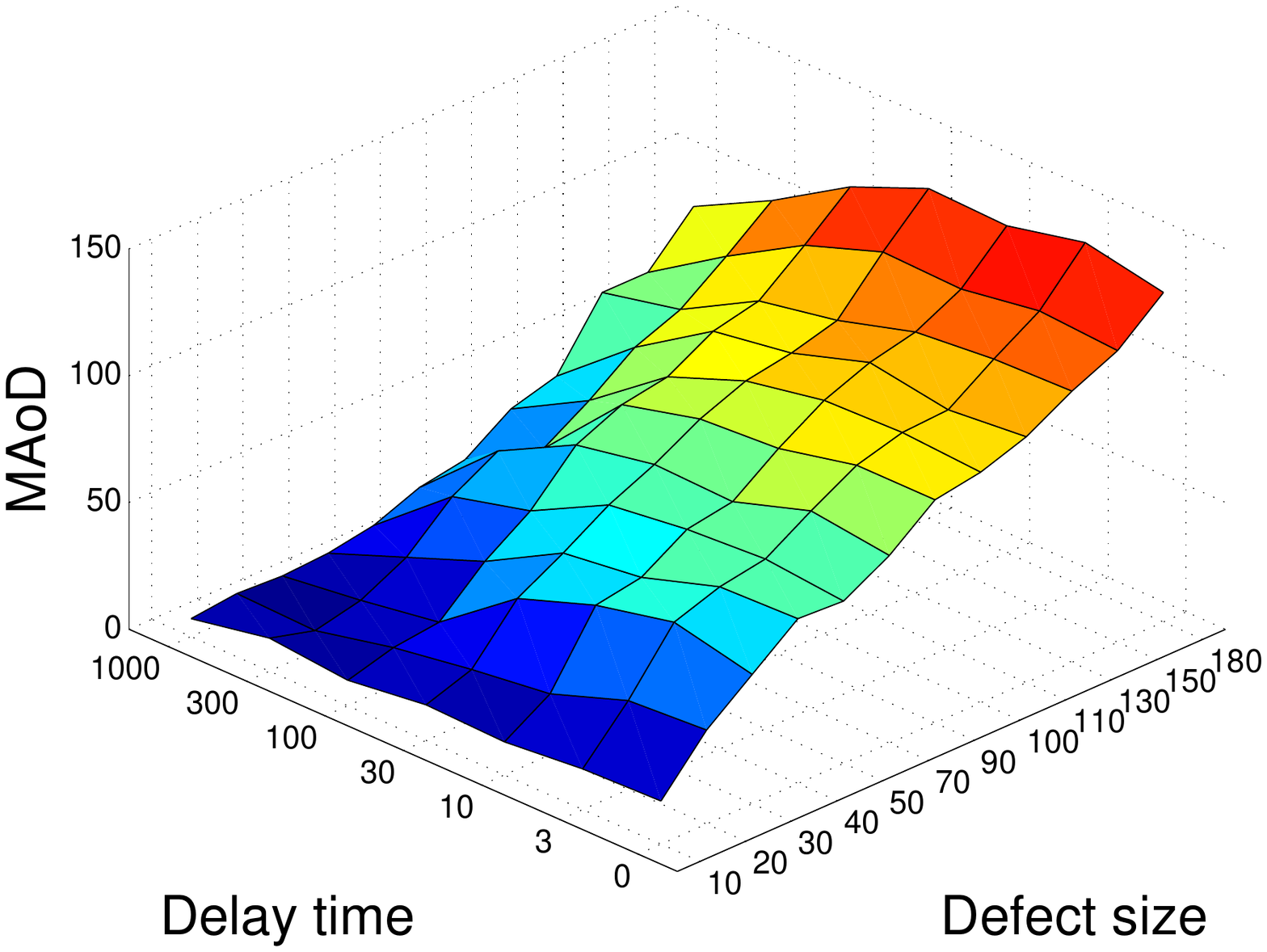}\label{l-gb-t-3d-s}}\\
\subfloat[]{\includegraphics[trim = 5mm 65mm 15mm 65mm, clip,width=0.4\textwidth]{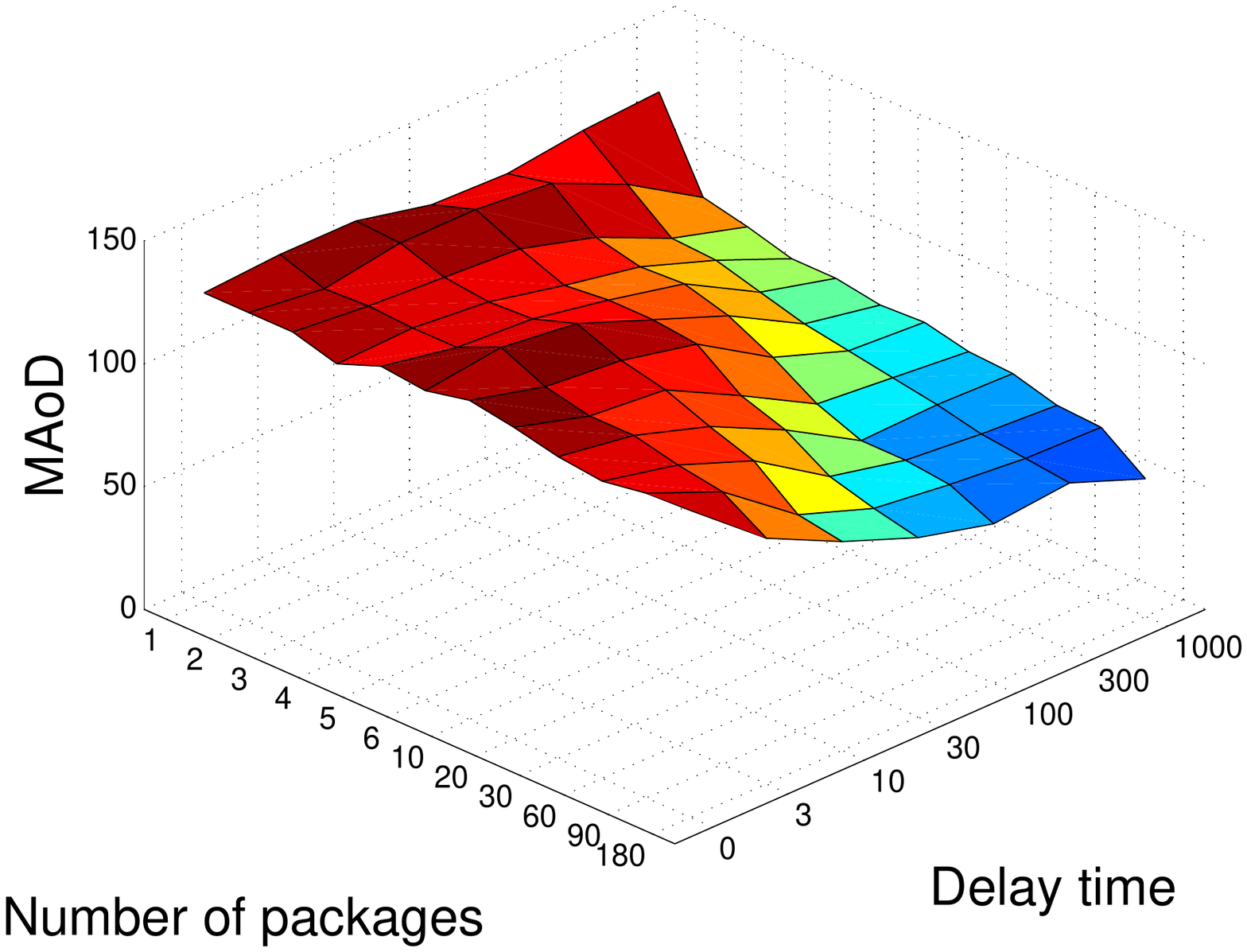}\label{l-b-s-3d-s}} 
   & \subfloat[]{\includegraphics[trim = 5mm 65mm 15mm 65mm, clip,width=0.4\textwidth]{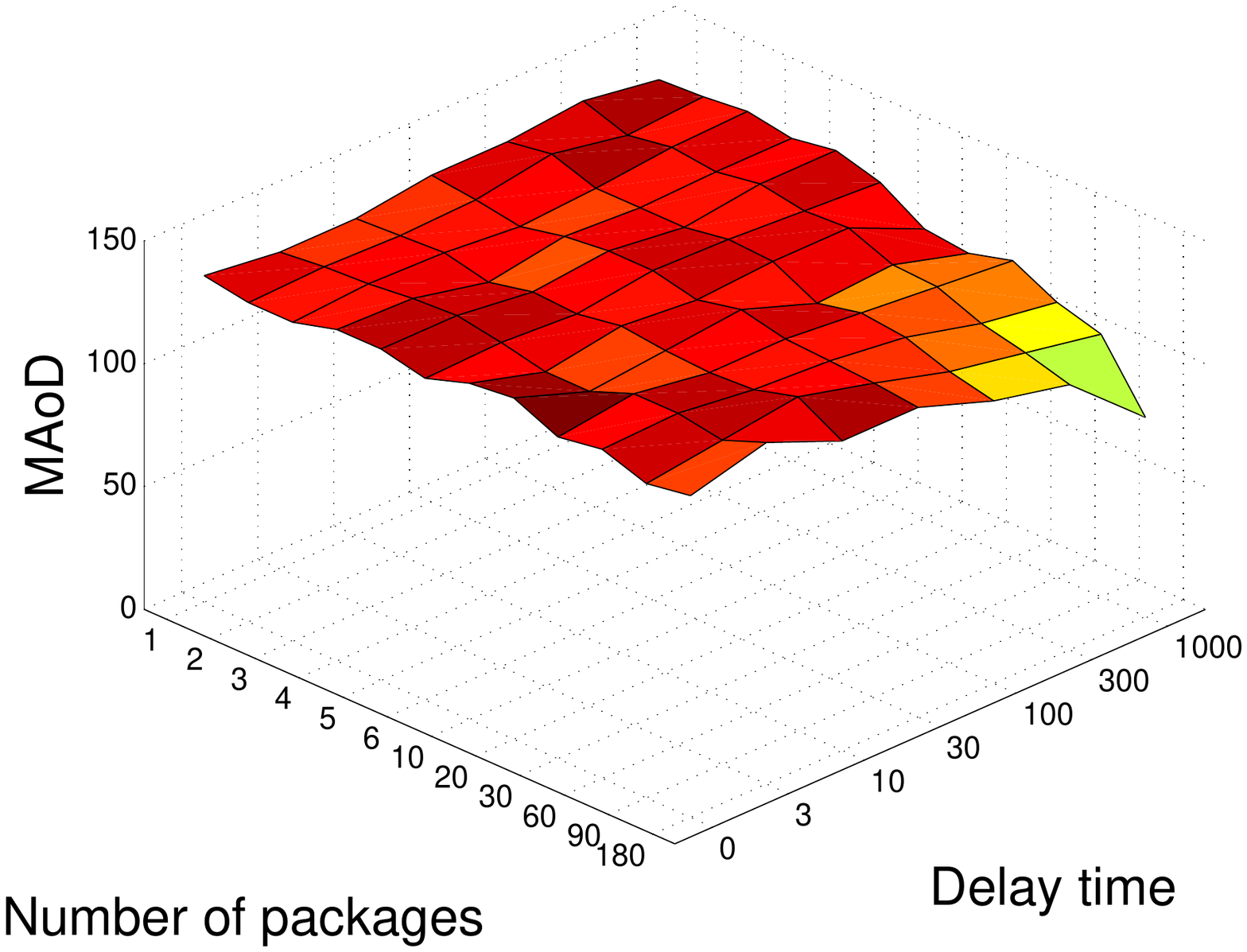}\label{l-b-t-3d-s}}\\
\end{tabular}
\end{center}
\caption{\footnotesize
\label{land-res-s}
The MAoD for different injury sizes and delays between removing nodes is plotted, using fixed (a) and modified $\eta$ (b). The same is plotted for different number of packages and delays between removing them using fixed (c) and modified $\eta$ (d). These diagrams are an average of nine runs with different initial conditions. Almost the same MAoD for different combinations of temporal and spatial pattern of injury.
   }\label{Mfigurelands}
\end{figure}

For better demonstration of the results, by omitting the complex behavior of each system we briefly find the maximum amount of disability which plays the main role in our study and plot this feature in figure \ref{l-gb-s-3d-s}. As it could seen, for large injury sizes and large delays the same MAoD "Maximum Amount of Injury" had been observed.

\subsubsection*{Type 2: No packages - Modified $\eta$}
In this section everything is like the previous section but the learning rate $\eta$ which is we used a modified $\eta$ as described in section \ref{modif-eta}. In the figure \ref{l-gb-s-t-res} the blue curves shows the behavior of this sub-model for some selected simulations.
The same as previous section The MAoD is plotted in figure \ref{l-gb-t-3d-s}. As it could seen, for large injury sizes and large delays the same MAoD "Maximum Amount of Disability" had been observed.
\subsubsection*{Type 3: packages - Fixed $\eta$}
Based on the finding and experiments of previous works, \cite{STEWART1951,Meyer1958,Glick1972,Rosen1971,Adametz1959,Finger1971,Patrissi1975} we know that removing the same amount of nodes in more than one package will decrease the MAoD too. To study this kind of injury in our model we set the injury size to $180$ and remove them in different number of packages. For example if the number of packages is equal to four and delay set to $30$, it means that every $30$ time steps we remove one quarter of the whole injury size ($45\,nodes$) together.
For the injury size of $180$ nodes and delays in the range of $0$, sudden injury, to $1000$ time steps, different systems had been simulated for $9$ different initialization. In the figure \ref{l-b-s-t-res} a selection of this sub-model has been shown in red.
The same as previous sections The MAoD is plotted in figure \ref{l-b-s-3d-s}.

\subsubsection*{Type 4: packages - Modified $\eta$}
For a fixed injury size of $180$ and modified $\eta$ described in \ref{modif-eta}, different systems had been simulated for various delay and number of packages.
In the figure \ref{l-b-s-t-res} a selection of simulation of this sub-model has been shown in blue.
The MAoD is plotted in figure \ref{l-b-t-3d-s}.



\section{Homeostasis model}
\label{Homeostasis model}
\subsection{The model}
In search for a simple and more biological model to work on we select a model based on Butz\cite{Butz2008,Butz2009}. This model is based on placing simple spiking neurons on a ring where 80\% of them are excitatory and the rest of them are inhibitory neurons. Neurons recive inputs on their dendrites and connect with other neuons through their axones. Here we have a total number of 100 neurons.
For more information on network topology see\cite{Butz2008,Butz2009}.

\begin{figure}[ht]
\begin{center}
\centerline{\includegraphics[height=7cm]{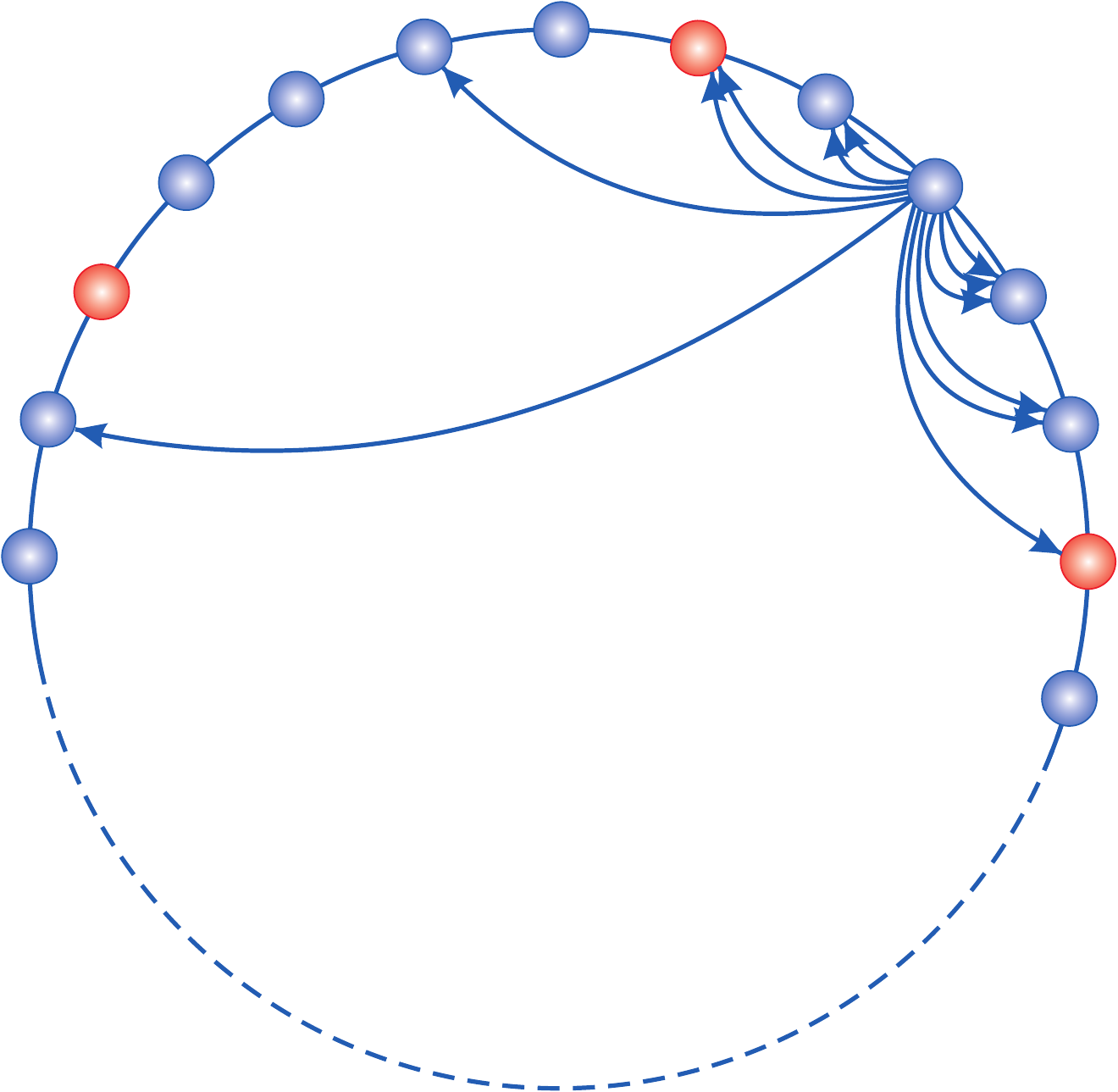}}
\protect
\caption{\footnotesize
\label{homeonet}
Homeostasis model network topology: 80\% of neurons (blue) are excitatory and the rest of them are inhibitory (red). Each neuron may has connections to its neighbors (Denderites and axones not shown). They can make synapse with nearer neighbors easier.
   }
\end{center}
\end{figure}

It has been shown \cite{Lipton1989,Mattson1988} that for biological networks there is a moderate firing rate that may be different for each network. In this model we set this to $0.5$ as we are not working on a specific network. The state of a network where the firing rate of neurons are similar to a moderate firing rate is called homeostasis.

There are two time steps in this model. One small time step called functional time step. In each functional time step we update the neuronal activity and the number of synapses dose not change. The second time step which is called morphological time step is the large time step. Here morphological time step is equal to 100 functional time steps. In each morphological time step according to neuronal activity in the last 1000 functional time steps we change the number of synapses in order to reach homeostasis.
In each functional time step first we calculate $X_i$ the presynaptic input of each neuron $i$. $X_i$ simply is a summation over all excitatory and inhibitory neurons which connected to neuron $i$ plus an external input which is from a poisson distribution. Then we update the firing probability $F$ of each neuron  with a sigmoid activation function.
After updating $F$ we update the state of each neuron. It may fire if a random number is smaller than $F$.
In each morphological time step we change the number of synapses according to the history of $F$. The aim of the network is to reach homeostasis so we demand that the average of $F$ over last 1000 functional time steps converges to $0.5$. So if it was grater than $0.5$ we change the number of synapses to decrease it. We do the opposite for dose neurons who have the average of $F$ smaller than $0.5$.
For axones, first we calculate
\begin{equation}\label{deltaa}
    \Delta A_i = \nu \cdot \Delta \overline{F_i} \cdot A_i
\end{equation}
where $\nu$ is a small number that adjust the rate of converging to homeostasis, $\Delta\overline{F_i} = \overline{F_i} - 0.5$ and $A_i$ is the number of connected axons of neuron $i$.
changing in the number of dendrites is similar to previous relation. You can see the detailed relations for change in synaptic elements in \cite{Butz2008,Butz2009}.
You can see in equation \ref{deltaa} that the changing rate is proportional to $\Delta \overline{F_i}$. $\Delta \overline{F_i}$ which shows how far is the neuron from the homeostasis can be interpreted as how bad this neuron works. By taking the variance of $F$ over all neurons we have a parameter which shows how far is the whole network from homeostasis.

The aim of this study is to see the relation between temporal patter of injury and network disability. In this model we suppose that the variance of $F$ over all nodes in the past 1000 functional time step shows resembles the network disability. To study the effect of temporal pattern of injury, first we let the network to reach homeostasis then we start killing the neurons in the network by disconnecting them from the other network nodes. For the same number of nodes to remove, one time we remove them all at once and one time gradually and one by one with a delay time of $"td"$ morphological time steps between each node removal.
The simulation details and parameter values for this part can be find in the Appendix \ref{homeoapp}.
\subsection{Homeostasis model resuls}
\begin{figure}[ht]
\begin{center}
\centerline{\includegraphics[height=7cm]{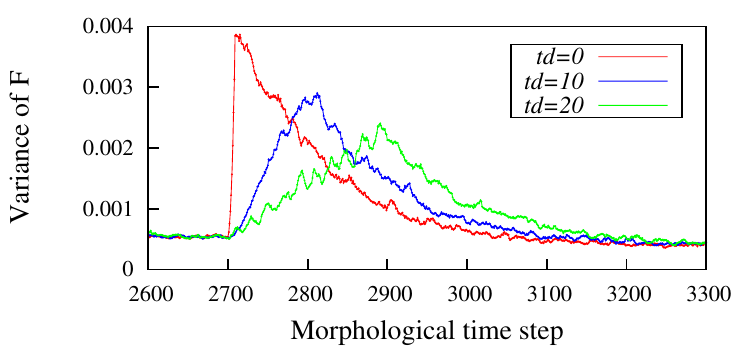}}
\protect
\caption{\footnotesize
\label{homeores}
The amount of network disability (shown by variance of $F$) for different delay time between node removal is shown. By increasing the delay time $"td"$ the maximum disability of network is decreased. Here blue line shows immediate defect but red and green ones show gradual defect with $10$ and $20$ time steps between node removal respectively. The maximum number of nodes removed in this diagram is $10$.
   }
\end{center}
\end{figure}

In figure \ref{homeores} the result for an injury size of $10$ is shown. In this figure the vertical axis shows the variance of $F$ which demonstrate the network disability and the horizontal axis shows morphological time step. Injury  starts at time $2700$ where the network reaches homeostasis. For the blue curve all the $10$ nodes are removed simultaneously but for red and green curves there are $10$ and $20$ morphological time steps respectively between removing each node. As clearly seen in the figure \ref{homeores} the amount of network disability lowers as the delay time between removing nodes is increased. This result is in agreement with the results in the three layer model.

One important thing about homeostasis model is that the recovery rate is proportional to the size of disability. That is the reason all three different injury patterns recover almost at the same time.

\section{Discussion}
Using two simple network models we studied the effect of temporal pattern of node destruction on network recovery. For both networks increasing the time delay between removing the nodes lead to a decrease in the maximum amount of network disability. In the case of our simple three layer model the amount of network disability is much higher when nodes in the process layer are removed at once (figure \ref{3layerres}). But when the node removal is more gradual the amount of disability is reduced. As shown in figure \ref{homeores} the same results holds for the more biologically inspired model,the homeostasis model. In both cases big sudden injury results in serious disability but large delay between removing nodes do not show serious disability. In general these results could be accounted for as a generic property of a distributed network where in the case of a gradual removal of nodes the remaining nodes take part in the process of recovery. More complex interactions,however, may also be involved explaining the difference of a sudden versus gradual damage. We studied the effect of learning rate dependence on the size of damage and indeed it is reported that following damage there is a time period in which the process of recovery speeds up \cite{Duffau2003}. Also, there are some animal models showing that a focal damage induces excitability and plasticity in the rest of the brain \cite{Buchkremer-Ratzmann1996}. Although the origins and mechanisms of these changes are still unknown it is worth studying these models in a more realistic setting, for example,it is known for a quite a while that brain recovery after sequential lesions depends on the amount of tissue resected at each surgical stage \cite{Stein1977}. In similar animal studies it is shown that in general a two-stage lesion has a much better chance of recovery than a one stage acute lesion \cite{Finger1971}. These are clear examples of what may be called a spatio-temporal pattern of brain damage. More importantly as has been argued by Daffauand et al. \cite{Duffau2003} the same sort of mechanism may be at work in aging up to a certain threshold where the system can no longer cope with the neural destruction. This indeed would be an interesting future research requiring a more realistic neural modeling.

\appendix
\section{Simulation details for three layer model}
\label{3layerapp}

In this simulation, One pattern which cover 50\% of the input layer teach to the system and we demand a patter in the output which covers 50\% of the output layer. 
For simulations with modified $\eta$, the value of $\eta$ for hamming distance below $h_0$ is equal to $\eta$ and for hamming distance above $h_1$ is rapidly decrease to $\eta_0$ (section \ref{modif-eta}).
The activation function for the middle layer is
\begin{equation}\label{activation2layer}
        F_i^{t}(X_i^{t})=\frac{1}{1+e^{-\frac{X_i^t - \theta_{ml}}{\beta}}}
\end{equation}
where $\theta_{ml}$ is threshold, $\beta$ is noise and $X_i^{t} = \sum_{j=1}^N w_{ij} \cdot j$ where $w_{ij}$ is the weight of connection from node $j$ in the input layer to node $i$ in the middle layer. Nodes in the output layer works on the basis of binary threshold model with threshold $\theta_{output}$.
In this model the learining rule is Hebb-like and the change in connection wights $\Delta w$ is:
\begin{equation}\label{change_w_in_3layer_model}
    \Delta w = \eta (a\cdot s_i \cdot s_j - b\cdot s_i - c\cdot s_j)
\end{equation}

where $a$, $b$ and $c$ are constants and $\eta$ is the learning rate, $s_i$ and $s_j$ show the state of nodes $i$ and $j$ which is $1$ if they fire and $0$ otherwise.

In this study all the constants set as shown in table \ref{tab:3layertable}.

\begin{table}[!h]
\begin{center}
\begin{tabular}{|l|l|l|}
\hline
Description & Symbol & Value \\
\hline
Total number of nodes & $N$ & 500\\
\hline
Subgroup of connected nodes & $n$ & 100\\
\hline
Middle layer threshold & $\theta_{ml}$ & 5.54\\
\hline
Output layer threshold & $\theta_{output}$ & 5.54\\
\hline
Noise & $\beta$ & 4\\
\hline
Hebbian learn parameter a & $a$ & 1.5\\
\hline
Hebbian learn parameter b & $b$ & 0.5\\
\hline
Hebbian learn parameter c & $c$ & 0.5\\
\hline
Maximum connection weight & $upperw$ & 10.3\\
\hline
Minimum connection weight & $lowerw$ & 0.0\\
\hline
Base learning rate & $\eta$ & 0.01\\
\hline
Begin of learning rate drop & $h_0$ & 5\\
\hline
End of learning rate drop & $h_1$ & 25\\
\hline
Decreased learning rate & $\eta_0$ & 0.001\\
\hline
\end{tabular}
\caption{Data used in three layer model simulations}
\label{tab:3layertable}
\end{center}
\end{table}

\section{Simulation details for homeostasis model}
\label{homeoapp}

In this model the activation function is
\begin{equation}\label{updatef}
    F_i^{t}(X_i^{t})=\frac{1}{1+e^{-\frac{X_i^t - \theta}{\beta}}}
\end{equation}
where $\theta$ is firing threshold and $\beta$ is determine noise by changing the steepness of the sigmoid function.

In this study all the constants set as shown in table \ref{tab:homeotable}.

\begin{table}[!h]
\begin{center}
\begin{tabular}{|l|l|l|}
\hline
Description & Symbol & Value \\
\hline
Total number of nodes & $length$ & 100\\
\hline
Morphological time step & $T$ & 100\\
\hline
Threshold & $\theta$ & 500\\
\hline
Noise & $\beta$ & 500\\
\hline
Probability of deletion & $pDeletion$ & 0.1\\
\hline
Learning rate & $\nu$ & 0.005\\
\hline
\end{tabular}
\caption{Data used in homeostasis model simulations}
\label{tab:homeotable}
\end{center}
\end{table}

You can see the more details for this model in [25, 26].

\bibliography{biblio}


\end{document}